\documentclass[%
 reprint,
 amsmath,amssymb,
 aps,
]{revtex4-2}

\usepackage{graphicx}
\usepackage{dcolumn}
\usepackage{bm}
\usepackage{hyperref}

\begin{document}

\preprint{APS/123-QED}

\title{High current ion beam formation with strongly inhomogeneous electrostatic field}

\author{S. S.~Vybin}
  \email{vybinss@ipfran.ru}
\author{I. V. Izotov}
\author{V. A. Skalyga}
    \affiliation{Institute of Applied Physics of Russian Academy of Sciences, 46 Ulyanov Str.,
603950, Nizhny Novgorod, Russia}


\begin{abstract}
A new approach to the development of extraction systems capable of forming ion beams with previously inaccessible intensity is proposed. The use of inhomogeneous accelerating field allows to improve the ion beam formation efficiency significantly. The increase of electric field magnitude is achieved by changing the shape of the electrodes only, without increasing the accelerating voltage and decreasing the interelectrode distance. The comparison is made between a new extraction system and a flat traditional one, which is the most common. The use of a new electrode geometry allows to increase the lifetime of the electrodes in sources of intense beams operating in a continuous wave mode. For electron cyclotron resonance ion sources, results demonstrate the possibility to form high-quality ion beams with a current density of more than $1 \, \text{A/cm}^2$.
\end{abstract}

\maketitle

\section{Introduction}

Intense ion beams are required in a wide range of basic and applied research. Beams of different energies and composition are used in the field of nuclear physics, medicine, materials science, surface treatment and modification, etc. \cite{lit1}. In most cases, further development of research requires an increase in current and a decrease in emittance of the beams. This ensures an increase in the brightness of the beam, which may be greatly affected by the ion beam extraction and formation system.

This paper proposes a new approach to the development of ion beam extraction systems. The main idea is to change the shape of the electrodes, which exploits an inhomogeneous distribution of the electric field in the accelerating gap of the ion beam formation system. This method may be especially effective in systems with a high current density, mitigating the negative effect of the space charge of the beam on its quality. It allows for significant enhancement of the current and/or reducing the emittance at a certain accelerating voltage and characteristic size of the system, especially when the extraction system operates in the space charge limited regime.

In the context of this article, only two-electrode systems are considered. The efficiency of the non-flat shape of the electrodes is demonstrated numerically with the IBSimu computational package \cite{lit2} (\url{http://ibsimu.sourceforge.net/}), which has proved its reliability in calculations of this type \cite{lit3, lit4}. As a starting point of the analysis of a new geometry of the extraction system, it is proposed to consider the combination of electrodes shown in Fig.~\ref{fig:fig1}a. Hereinafter we denote such a system as a “spherical” because the puller and the tip of the plasma electrode form plates of a hollow hemispherical capacitor (this region is highlighted with a rectangle in Fig.~\ref{fig:fig1}a).

The authors suggest the use of such extraction system in high-current ECR sources like the SMIS 37/75 \cite{lit5} and GISMO \cite{lit6} with plasma heated by a high power ($10-200$ kW) millimeter-wave radiation of gyrotrons with a frequency of $28 - 75$ GHz. All calculations are performed for the plasma parameters and the geometry of the extraction system typical for these facilities. For the sake of clearness, all calculations in the work are carried out for ion beams with atomic hydrogen dominating the composition. Despite the fact that we concentrate on ECR ion sources, we emphasize that the results presented in this work are universal and do not depend on the method of plasma formation and the type of ions in the formed ion beam.
\section{Advantages of the new extractor}
\label{sec:sec2}

\begin{figure*}
\includegraphics{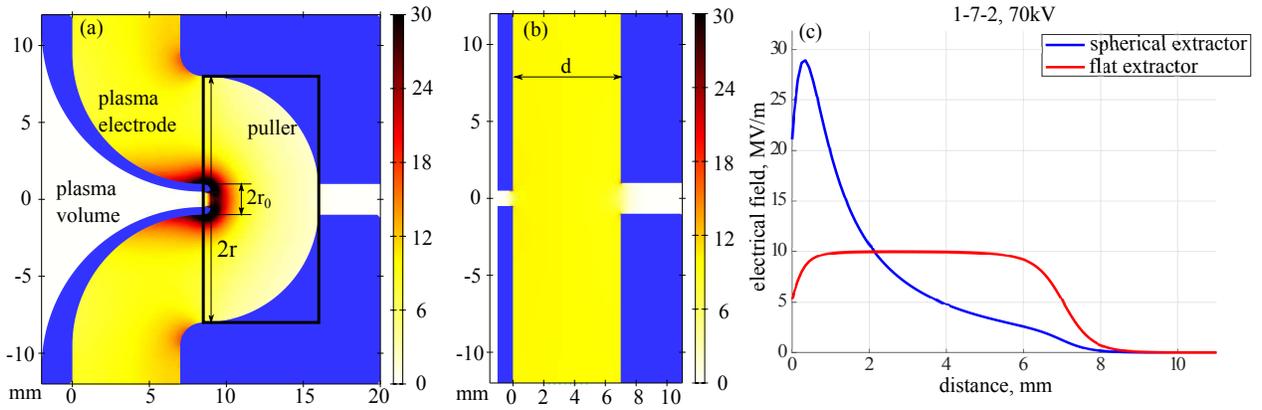}
\centering
\caption{\label{fig:fig1} Distribution of the electric field modulus (MV/m) in the interelectrode space (a - spherical, b - flat). Distribution of the electric field along the axis of symmetry (c). 
The voltage is $70$ kV. The origin on the graph (c) is aligned with the edge of the plasma electrode. The use of a new geometry allows to increase the electric field magnitude in several times. The radii of curvature of the plasma electrode and puller are denoted by $r_0$ and $r$, respectively, the distance between electrodes $d = r - r_0$. The new extractor maintains a constant distance between the electrodes. It should be noted that the radius of the hole of the plasma electrode is less than the radius of its curvature.
}
\end{figure*}

The problem of the dense ion beam formation is due to the harmful influence of the self space charge, which distorts the shape of the meniscus, making it convex. Accordingly, the accelerating field near the meniscus forms a diverging beam, which cannot be transported efficiently. The increase in the electric field near the meniscus provides higher beam acceleration gradient and reduces the region with high space charge level. 
The use of inhomogeneous accelerating field allows an increase of the electric field magnitude near the plasma meniscus without raising the extraction voltage or reduction of interelectrode distance (see Fig.~\ref{fig:fig1}c.). In this work we consider a new electrode geometry which produces an inhomogeneous electric field providing higher gradients of ion beam acceleration and reducing the space charge influence on its quality. 

The effectiveness of the new geometry is compared to a flat single-aperture one. The essential parameters are the distance between electrodes and the apertures. We name the set of these parameters as a configuration and denote for brevity as $D_1-L-D_2$, where $D_{1,2}$ are the apertures of the plasma and puller electrodes, respectively (in mm), $L$ is the interelectrode distance (in mm).

The new extractor creates a region of strong electric field in the plasma electrode tip area (see Fig.~\ref{fig:fig1}a). This fact makes it possible to increase the maximum current density of the formed ion beam. Let us explain this result from the point of view of the Child-Langmuir law. The values of the space charge-limited current densities of beams formed by an “ideal” extractors with flat and spherical symmetries are expressed as follows \cite{lit7}.

Flat case:
\begin{eqnarray*}
    J_{fl} = \frac{4\varepsilon_0}{9}\sqrt{\frac{2e}{m}}\frac{U^{3/2}}{d^2}.
\end{eqnarray*}

Spherical case:
\begin{eqnarray*}
    J_{sph} = \frac{4\varepsilon_0}{9}\sqrt{\frac{2e}{m}}\frac{U^{3/2}}{r_0^2\alpha^2(r/r_0)}.
\end{eqnarray*}

The following notation is used in the formulas shown above: $U$ is the accelerating voltage, $J$ is the current density, $e$ is the ion charge, $m$ is the mass of the ion, $\varepsilon_0$ is the dielectric constant, $d$ is the distance between the electrodes in the planar case, $r$ is the collector radius, $r_0$ is the emitter radius, see Fig.~\ref{fig:fig1}, $\alpha(r/r_0)$ is the Langmuir-Blodgett function for the spherical case.

\begin{figure}[b]
    \includegraphics{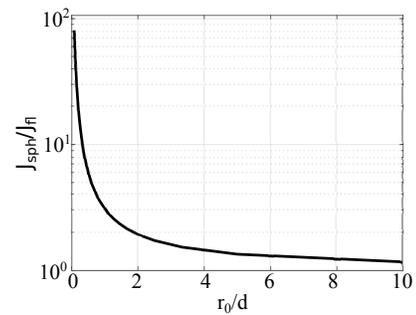}
    \caption{\label{fig:fig2} The dependence of the maximum current density ratio in accordance with the Child-Langmuir law on the radius of the plasma electrode normalized to the distance between electrodes.}
\end{figure}

The Child-Langmuir law also allows one to determine general patterns in the behavior of the maximum beam current density depending on the geometry parameters (see Fig.~\ref{fig:fig2}).

As it is seen from  Fig.~\ref{fig:fig2}, with a decrease in the emitter radius, an increase in the current density limit is observed in the  spherical case. This is due to the fact that, in a spherical capacitor, the field on the surface of the inner electrode is proportional to $E \propto 1/r_0^2$. A flat extractor can be represented as spherical with a radius of the emitter tending to infinity. Therefore, its current density limit does not depend on the emitter radius. As the diameter of the emitter increases, the performance of the spherical extractor levels with a flat one. Thus, the limit of the applicability of the spherical extractor may be expressed as $d \ll r_0$. Despite the fact that, as it follows from Langmuir's work, a spherically symmetric extractor is the most effective for high-density beams, the authors were not able to find examples of its real applications for the formation of high-current ion beams in the literature.

\begin{figure}[b]
   \includegraphics{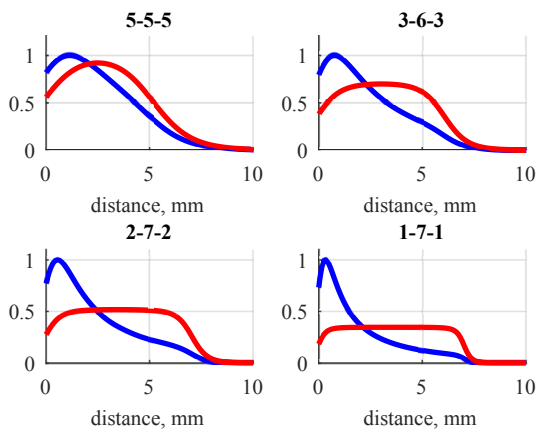}
    \caption{\label{fig:fig3} Graphs of the distribution of the electric field for different configurations of the extractor. The electric field is normalized to the maximum value for each case. The spherical geometry is blue and the flat is red.}
\end{figure}

Naturally, it is impossible to use the truly spherically symmetric geometry in an ECR ion source, as it is normally  based on open magnetic traps (i.e. simple mirror trap \cite{lit8} or traditional “min-B” ECRIS scheme \cite{lit9}). Therefore, the proposed extractor is not strictly spherical. However, it is similar to ideal systems as it redistributes the electric field in between electrodes. Consequently, the aperture of the plasma electrode plays a significant role in the ability of the extraction system to form dense ion beams. The maximum efficiency of the new extractor is achieved when $d \gg r_0$, when compared to a flat one. To illustrate this fact, a comparison of the distribution of the electric field in the flat and spherical case for various configurations is shown in Fig.~\ref{fig:fig3}.

A new approach to the design of extraction systems opens up a new way to increase the electric field near the emitting surface. In the traditional (flat) geometry, this may be achieved  with the increase of the extraction voltage and/or the decrease of the interelectrode distance only, while the electric field in the accelerating gap remains uniformly distributed, and the maximum electric field is limited by the electrical breakdown of the interelectrode space. When using the proposed geometry, the electric field increases only near the plasma electrode, and the average field on the puller surface remains low, thus  reducing the probability of breakdown, which makes it possible to further increase the extraction voltage.
\section{Comparison with flat electrodes}

The process of the ion beam extraction and formation was modelled using the IBSimu package with the following plasma parameters, conventional for SMIS 37 ion source: the plasma potential is $10$ V, the ion temperature is $1$ eV and the electron temperature is $10$ eV. The configuration of the extractors is selected $3-10-4$ for both flat and spherical cases. Accelerating voltages equal to $40$, $60$, and $80$ kV were simulated. For an interelectrode distance of $10$ mm, the maximum reliable extraction voltage is $80$ kV, according to \cite{lit10}. Thus, decrease of the extraction voltage reduces the risk of electrical breakdown. The magnetic field is not accounted in simulation due to the fact that the new approach in the extraction system design can be applied to a wide variety of ion sources with different magnetic structures. 

\begin{figure}[h!]
    \includegraphics{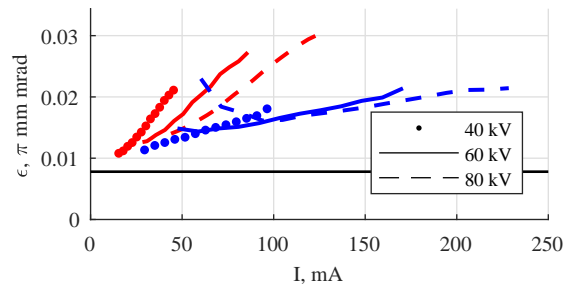}
    \centering
    \caption{\label{fig:fig4} The dependence of the normalized RMS emittance of the beam on the total extracted current. The flat case is red and the spherical case is blue. The black horizontal line is the emittance of the $\text{H}^+$ beam with temperature $T = 1$ eV and diameter $D_1 = 3$ mm.}
\end{figure}

One of the most important characteristics of an ion beam for accelerator application is its transverse emittance. The dependence of the normalized RMS emittance on the total beam current is shown in Fig~\ref{fig:fig4}. In spherical geometry, a lower emittance is achieved at the same current level as in the flat case. This is due to the lower space charge influence on the beam in the spherical case. The similar emittance as in the spherical geometry may be achieved in the flat geometry only by means of a significant increase in the voltage.

When the ion beam is extracted in the optimal mode for the flat geometry, a substantial decrease in the extraction voltage is possible using the spherical geometry.

\begin{figure}[h!]
   \includegraphics{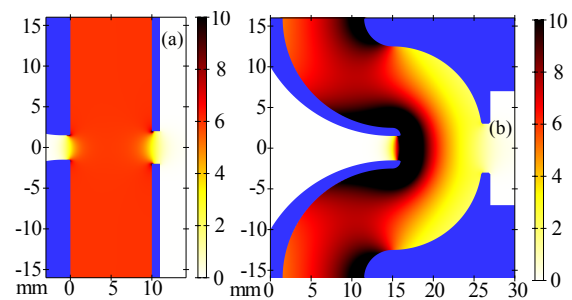}
   \centering
    \caption{\label{fig:fig5} The distribution of the electric field (MV/m) in a planar extractor (colormap) with a potential difference of $60$ kV (a) and spherical at $80$ kV (b). It is possible to increase the value of the optimal current by $54\%$ in accordance with the Child-Langmuir law. The electric field on the puller at $R=\sqrt{X^2+Y^2}<10$ mm is $3$ times less than in the flat case.
}
\end{figure}

In addition, the magnitude of the electric field on the puller close to the beam envelope decreases in the spherical geometry when compared to the flat one. It allows us to further expand the range of extraction voltages used (see Fig.~\ref{fig:fig5}). Since the extraction systems presented in this article have cylindrical symmetry, we consider the $Z$ axis to be the longitudinal axis of symmetry, whereas $X$ and $Y$ - transversal ones. The origin and orientation of the transverse axes is not essential.
\section{The effect of the puller aperture on the ion beam extraction}

\begin{figure}[h!]
   \includegraphics{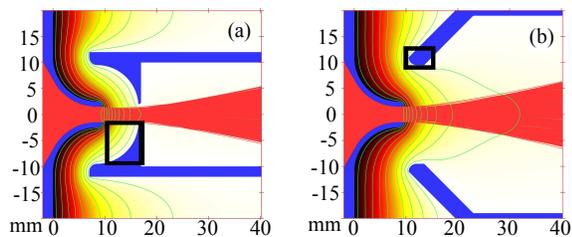}
   \centering
    \caption{\label{fig:fig6} Formation of an ion beam with various puller designs. A significant increase in the aperture of the puller slightly affects the shape of the meniscus and the mode of extraction of the beam. The puller part highlighted in the figure (a) has a weak effect, while the highlighted part in the figure (b) makes the main contribution.}
\end{figure}

As it follows from the simulations, with an increase in the puller aperture, there is no significant deterioration in the quality of the formed beam. Fig.~\ref{fig:fig6} demonstrates the ion trajectories and the electric field distribution in the spherical geometry with significantly different puller apertures. As it is seen, the puller aperture can be significantly expanded, while the shape of the meniscus is barely affected, and so are the ion trajectories.

This is due to the fact that the formation of a strong field region in the proximity of the plasma electrode tip is mainly governed by the part of the puller highlighted in Fig.~\ref{fig:fig6}b (and the plasma electrode shape). The part of the puller highlighted in Fig.~\ref{fig:fig6}a is responsible for the distribution of the electric field in the region, where the ion beam has already gained a significant part of its total energy, thus having minor influence on its trajectories. The use of the spherical geometry enables a significant increase in the puller aperture. This ensures the absence of collisions of the ions with the puller, eliminating the secondary electron flux and prolonging the lifetime of the puller electrode, which may be an issue in a CW operation mode of an ion source.
\section{Prospects for improving the ion beam parameters at the SMIS 37 facility}

The record results on the formation of proton beams at the SMIS 37 facility were reported in \cite{lit11}. Proton beam currents of up to $500$ mA were achieved in experiments with the plasma electrode aperture of $1$ cm.

\begin{figure}[t]
   \includegraphics{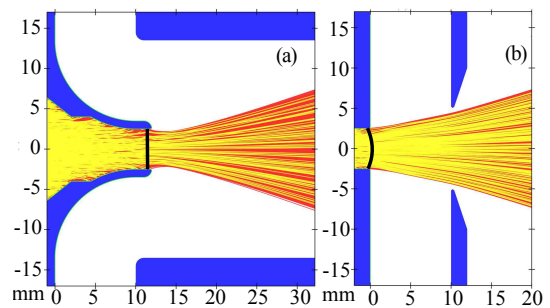}
    \caption{\label{fig:fig7} Calculation of beam formation with the following parameters. Beam content: $94\%\, \text{H}^+$, $6\%\, \text{H}_2^+$, ion temperature: $1$ eV, electron temperature: $10$ eV, plasma potential: $50$ V, extraction voltage: $50$ kV. Average current density: $1.31 \, \text{A/cm}^2$ for the planar case and $1.75 \, \text{A/cm}^2$ for spherical geometry. The magnetic field strength in the extraction region is $0.33$ T. The aperture of the plasma electrode is $5$ mm, the interelectrode distance is $10$ mm. The total current is $257$ mA for the flat case (a) and $347$ mA for the spherical case (b), while the normalized RMS beam emittances turn out to be approximately the same and equal to $0.12$ $\pi$ mm mrad. The black curve denotes the meniscus shape. 
    }
\end{figure}

The use of a spherical extraction system with an aperture of a plasma electrode of $10$ mm at the interelectrode distance of the same level is inefficient. It is necessary to exceed the interelectrode distance over the aperture of the plasma electrode by several times. Therefore, the required minimum interelectrode distance is about $20$ mm. It forces the use of a higher extraction voltage, which could not be achieved at SMIS 37 facility. Thus, hereafter we consider  the plasma electrode aperture of $5$ mm. Then, with an interelectrode distance of $10$ mm, a spherical extractor can show its advantage over the flat one at available voltage levels. The simulation results corresponding to experimental conditions with a plasma electrode aperture of $5$ mm are presented in Fig.~\ref{fig:fig7}. The increase of the extracted current in the spherical case is achieved with raising the plasma emissivity until the beam trajectories became roughly the same as in the flat case. In the experiment, a lower emittance value of $0.07$ $\pi$ mm mrad was measured than in the simulations \cite{lit11}. The origin of the difference is believed to be the influence of a noticeable space charge compensation in the experiment, whereas the calculations were carried out without compensation. Thus, the $35\%$ increase in the total current is observed for the spherical geometry over the flat one. 

It is of note that beam parameters acquired in the spherical case exceed the requirements of modern ambitious accelerator projects, where proton injection is required or is seen as an attractive option if $> 200$ mA $\text{H}^+$ current with low emittance ($< 0.1$ $\pi$ mm mrad) can be demonstrated. Examples of such projects include DARIA (a compact LINAC-based neutron source, beam requirements: $\text{H}^+$, $100$ mA, $< 0.1$ $\pi$ mm mrad \cite{lit12}) and the ISIS-II upgrade in the UK where short pulse proton injection is considered in parallel to charge exchange injection of $\text{H}^-$ into a rapid-cycling synchrotron, i.e. ISIS-II upgrade (private communications).
\section{Conclusions}

A new approach to the design of extraction systems can significantly increase the intensity of the ion beam. This may be achieved by the change of the electrodes shape only. The formation of intense proton beams with a current density of $1.75$ $\text{A/cm}^2$ is shown numerically with a possibility to further increase up to tens of $\text{A/cm}^2$, current density being limited by the capabilities (i.e. the plasma emissivity) of the plasma generator only. The possibility of a significant increase in the aperture of the puller electrode provided by the new system may be beneficial from an engineering point of view. 

The proposed geometry might be of interest especially for the ion sources operating in a space-charge-limited mode. In particular, the spherical extractor may be valuable for $\text{H}^-$ beam formation, when the increase in the total beam current is desired without raising the extraction voltage. 

However, the new approach possesses several flaws. A flat extractor allows the formation of an ion beam with a smaller divergence than a spherical one. This is due to the fact that a spherical extractor has a non-negligible imminent transverse component of the electric field. Nonetheless a larger divergence  may be compensated by a smaller beam size provided by the new system. The prospect of using the proposed electrodes gives additional motivation in the study of intense plasma flux production.
\begin{acknowledgments}
The work was supported by the project of the Russian Science Foundation Grant No. 16-19-10501. We thank Dr. Olli Tarvainen for fruitful discussions and valuable comments.
\end{acknowledgments}


\bibliography{main}

\end{document}